\documentclass{elsartmy}
\usepackage[koi8-u]{inputenc}
\usepackage[ukrainian]{babel}

\begin{document}

\begin{frontmatter}

\title{%
Estimation of Possible Selectivity and Sensitivity of a
Cooperative System to Low-Intensive Microwave
Radiation
}
\thanks{Talk made on the
"Electromagnetic Hypersensitivity, 2nd Copenhagen Conference", Copenhagen,
 1995}

\author{A.K.Vidybida}

\address{Bogolyubov Institute for Theoretical Physics\\
Kyiv 03143, Ukraine\\E-mail: vidybida@bitp.kiev.ua}

\end{frontmatter}
\section{}
There are many cases of extremely high sensitivity to low-level
factors. In some cases the low-level factors may be benefitial. For
example homeopathic pills, or manual therapy. In some cases they may
be harmful, such as technogenic environmental factors. Other cases
represent a very sensitive communication. For example, communication
by odorants between some butterflies, or cases of so called
extrasensory perception.

\section{}
It is interesting that high sensitivity is mostly observed for the living
objects. The living objects are open systems. That means they
exchange substances with the external world.

All living systems, even a single cell, possess a sort of free will.
Speaking in physical language, they are multistable. The
multistability means that system can have several
different internal states,
while the external conditions remain the same. In
other words, the external conditions cannot determine uniquely the
internal state of a living object.
It is namely this fact that is treated as a
sort of free will.

\section{}
The effect of microwaves on living objects is studied experimentally for
more then 20 years \cite{1,2}. During
the last 10 years microwaves are used for
healing
\cite{3,4}. The microwaves traditionally belonge to physical factors and
their effects have to be reduced to purely physical mechanisms. Till now
such mechanisms has not been found. The problem is that microwaves are
applied in a very low doses. At the level of a single organic molecule
this low-intensive microwaves produce effect which is well jammed by
thermal oscillations of the molecules. The common physics usually
deals with gases, or liquids, or solids, and it has nothing to say in
our situation. Our opinion is that the answer may be found in the
internal complexity of the object the microwaves are acting upon.

\section{}
This paper is aimed to explain how the weak microwave radiation may
produce a pronounced effect in a coherent system comprizing a large
number of molecules. The following model chemical system is
considered.

\def\cas{$C^*$}
\def\ctil{$C~\tilde{ }$ }

\begin{equation}
A + 2X \rightleftharpoons 3X
\end{equation}

\begin{equation}
B + X \rightleftharpoons C^*
\end{equation}

\begin{equation}
C^* \rightleftharpoons C~\tilde{ }
\end{equation}
\bigskip\bigskip

Here $X$, $A$, $B$, $C$ denote molecules of different species.
\cas and \ctil denote active and non-active states of the same
molecule $C$.  The $A$, $B$, and $C$ concentrations are maintained
constant. Concentration of $X$ is expected to be established
spontaneously via the chemical reactions noted here.

This system is open. Indeed, as $A$, $B$, $C$ can be produced or
consumed, the chemicals have to be removed or added to the system in order to
keep constant concentrations.

This system is multistable. It is established mathematically that
under fixed conditions two different concentrations of $X$, are
possible as its stable internal state: $[X]=x_1$, or $[X]=x_2,\quad
x_1<x_2$. Both $x_1$ and $x_2$ are self-sustaining due to positive
and negative cooperativity presence in reaction (1). Switching
between $x_1$ and $x_2$ states may happen due to external
influencees, or internal fluctuation processes.

\section{}
If the reactor has a big volume, two stable concentrations may coexist
in such a reactor, occupaing different parts of the volume. We
consider the small enough reactor in wich only single stable
concentration is possible. This is a coherent reactor. The coherent
reactor responses to external influences as a single unit, namely, it
switches from one stable state to another one in the whole volume
at onse. The number of $C$-type molecules in the coherent volume is
essential for sensitivity: the more molecules, the more sensitive the
system is. This number is estimated for biologically realistic
parameters as $N\sim 10^9 \div 10^{12}$.

\section{}
It is known that microwaves are able to shift equilibrium in some
chemical reactions, but the shift may be very small. This is because
molecular vibrations in this frequency band is damped by viscous
friction. If the mw radiation power surfase density is
$I=1$ mW/cm$^2$
one may expect the concentration increment
\begin{equation}
{\Delta [C^*]\over
[C^*]} \sim 10^{-7}.
\end{equation}
As we know, the system (1)--(3) is able to switch from its stable
state to another one due to fluctuations. This process is
characterized by a mean waiting time for first switching,
$T_{x_1\to x_2}$. If an external influence is added, the mean waiting
time will change. How much it changes depends on system sensitivity.
Mathematical treatment under some simplifying assumptions
(for detailes see \cite{5,6}) gives the
figure:
\begin{equation}
 {T({\rm without\ field}) \over T({\rm with\ 1\ mW/cm}^2
{\rm\ field})} \sim \exp (10^4)
 \end{equation}

Comparing (4) and (5), one may conclude that single receptor molecules
are very little influenced by microwaves. But if a large number of
such receptors are incorporated within a living-like multistable
coherent system, the whole system may react very sharply.

\section{}
In connection with this conclusion there are two interesting
observations:

The first one is the reductionism problem \cite{7}. If we have here such a
sensitive system, is it possible to reduce its properties to the
properties of its parts. Theoretical basis for such reduction has
been
offered in this paper. But what may be seen in experiments? The $C$-type
molecules realizing primary reception of microwaves show extremely
small effect, whereas whole system manifests a very high effect. It
seems incredible that the effects are in causal relation.

The second observation is that the model system that we have
considered, has at least two common features with usual sensory
systems, such as olfaction, taste and others \cite{8}. Namely, in analogy with
sensory systems this system is hierarchical. The first hierarchical
level here is the $C$-type molecules, and the second one is the level
of the whole system, including the cooperatively coupled $X$-molecules.
Also, in analogy with known sensory systems, we have here a
threshold-type behavior: a definite threshold must be overcome for
switching.

\end{document}